 \documentstyle[emulateapj,apjfonts,psfig]{article}

\newcommand{\hMpc}{$h^{-1}{\ }{\rm Mpc}$}
\newcommand{\hMsun}{$h^{-1}{\ }{\rm M_{\odot}}$}

\newcommand{\Dvir}{$\Delta_{\rm vir}$}

\newcommand{\Rhalo}{$R$}

\newcommand{\Mhalo}{$M$}

\newcommand{\Table}[1]{Table~\ref{#1}}

\newcommand{\Eq}[1]{Eq.~(\ref{#1})}
\newcommand{\Fig}[1]{Fig.~\ref{#1}}

\newcommand{\ea}{et~al.~}                            

\newcommand{\astroph}[1]{\mbox{\texttt{astro-ph/#1}}}

\begin{document}

\title{On the Correlation between Spin Parameter and Halo Mass}

\author{Alexander Knebe\altaffilmark{1}, 
        Chris Power\altaffilmark{2}}

\altaffiltext{1}{Astrophysikalisches Institut Potsdam,
				An der Sternwarte 16,
				14482 Potsdam,
				Germany}
\altaffiltext{2}{Centre for Astrophysics \& Supercomputing,
				Swinburne University,
				Mail H31, PO Box 218,
				Hawthorn, VIC 3122,
				Australia}

\begin{abstract}
  We report on a correlation between virial mass $M$ and spin parameter
  $\lambda$ for dark matter halos forming at redshifts $z \gtrsim 10$.
  We find that the spin parameter decreases with increasing halo mass.
  Interestingly, our analysis indicates that halos forming at later
  times do not exhibit such a strong correlation, in agreement with 
  the findings of previous studies. We briefly discuss the implications 
  of this correlation for galaxy formation at high redshifts and the galaxy
  population we observe today.
\end{abstract}


\section{Introduction} \label{sec:introduction}

The physical mechanism by which galaxies acquire their angular momentum
is an important problem that has been the subject of investigation for
nearly sixty years (Hoyle 1949). This reflects the fundamental role played
by angular momentum of galactic material in defining the size and 
shapes of galaxies (e.g. Fall \& Efstathiou 1981). Yet despite its
physical significance, a precise and accurate understanding of the
origin of galactic angular momentum remains one of the missing pieces 
in the galaxy formation puzzle.

A fundamental assumption in current galaxy formation models is that
galaxies form when gas cools and condenses within the potential wells
of dark matter halos (White \& Rees 1978). Consequently it is probable that 
the angular momentum of the galaxy will be linked to the angular
momentum of its dark matter halo (e.g. Fall \& Efstathiou 1980; Mo, Mao
\& White 1998; Zavala, Okamoto \& Frenk 2007). Within the context
of hierarchical structure formation models, the angular momentum growth 
of a dark matter proto-halo is driven by gravitational tidal torquing during 
the early stages (i.e. the linear regime) of its assembly. This ``Tidal Torque
Theory'' has been explored in detail; it is a well-developed analytic 
theory (e.g. Peebles 1969, Doroshkevich 1979, White 1984) and its 
predictions are in good agreement with the results of cosmological 
$N$-body simulations (e.g. Barnes \& Efstathiou 1987; Warren~\ea 1992; 
Sugerman, Summers \& Kamionkowski 2000; Porciani, Dekel \& Hoffman
2002). However, once the proto-halo has passed through maximum
expansion and the collapse has become non-linear, tidal torquing
no longer provides an adequate description of the evolution of the 
angular momentum (White 1984), which tends to decrease with time. 
During this phase it is likely that merger and accretion 
events play an increasingly important role in determining both the
magnitude and direction of the angular momentum of a galaxy (e.g. Bailin \&
Steinmetz 2005). Indeed, a number of studies have argued that mergers
and accretion events are the primary determinants of the angular
momenta of galaxies at the present day (Gardner 2001; Maller, Dekel~\&
Somerville 2002; Vitvitska~\ea 2002).\\

It is common practice to quantify the angular momentum of a dark matter 
halo by the dimensionless ``classical'' spin parameter (Peebles 1969),
\begin{equation}
  \label{eq:lambda}
  \lambda = \frac{J \sqrt{|E|}}{GM^{5/2}},
\end{equation}
where $J$ is the magnitude of the angular momentum of material within
the virial radius, $M$ is the virial mass, and $E$ is the total energy 
of the system. It has been shown that halos that have suffered a recent 
major merger will tend to have a higher spin parameter $\lambda$ than 
the average (e.g. Hetznecker \& Burkert 2006; Power, Knebe \& Knollmann 
2008). Therefore one could argue that within the framework of hierarchical
structure formation that higher mass halos should have larger spin 
parameters \emph{on average} than less massive systems because they 
have assembled a larger fraction of their mass (by merging) more
recently. 

However, if we consider only halos in virial equilibrium, should we
expect to see a correlation between halo mass and spin? One might
na\"ively expect that more massive systems will have had their maximum
expansion more recently and so these systems will have been tidally
torqued for longer than systems that had their maximum expansion at
earlier times. This suggests that spin should \emph{increase} with
timing of maximum expansion and therefore halo mass. However, one
finds at best a weak correlation between mass and spin for equilibrium
halos at $z$=0 (e.g. Cole~\& Lacey 1996; Maccio et al. 2007; Bett et
al. 2007, hereafter B07), and the correlation is for spin to
\emph{decrease} with increasing halo mass, contrary to our na\"ive
expectation.

In this paper, we report on a (weak) correlation between spin and
mass for equilibrium halos at redshift $z$=10. The trend is for 
higher-mass halos to have smaller spins, and is qualitatively
similar to the one reported by B07 for the halo
population at $z$=0. We present the main evidence in support of this
correlation in Section~\ref{sec:correlation} and we consider its
implications for galaxy formation in Section~\ref{sec:conclusions}.

\section{The Simulations} \label{sec:simulations}


For the simulations presented in this paper we have adopted the
cosmology as given by Spergel~\ea (2003) ($\Omega_0=0.3$,
$\Omega_{\Lambda}=0.7$, $\sigma_8=0.9$, and $H_0=70$km/sec/Mpc). Each
run employed $N=256^3$ particles and differed in simulation box-size
$L_{\rm box}$, which leads to the particle mass $m_p$ differing
between runs -- $m_p=\rho_{\rm crit} \Omega_0 (L_{\rm box}/N)^3$,
where $\rho_{\rm crit}=3H_0^2/8\pi\,G$. This allows us to probe a
range of halo masses at redshift $z$=10. The primary parameters of
these simulations are summarized in \Table{tab:simu}.

Halos in all runs have been identified using the MPI parallelized
version of the \texttt{AHF} halo finder\footnote{\texttt{AHF} is
  already freely available from
  \texttt{http://www.aip.de/People/aknebe}}
(\texttt{AMIGA}'s-Halo-Finder), which is based on the \texttt{MHF}
halo finder of Gill, Knebe \& Gibson (2004). For each halo we compute
the virial radius \Rhalo, defined as the radius at which the mean
interior density is \Dvir\ times the background density of the
Universe at that redshift. This leads to the following definition for
the virial mass \Mhalo:

\begin{equation} \label{eq:Mvir}
  M = \frac{4\pi}{3} \Delta_{\rm vir} \Omega \rho_{\rm crit} R^3 \ .
\end{equation}

\noindent Note that \Dvir\ is a function of redshift and amounts to
$\Delta_{\rm vir} \approx 210$ at redshift $z$=10 , $\Delta_{\rm vir}
\approx 230$ at $z$=1, and the ``usual'' $\Delta_{\rm vir} \approx
340$ at $z$=0 (cf. Gross 1997). \Table{tab:simu} summarises the total
number of halos ($N_{\rm halos}$) recovered by \texttt{AHF}, while
$z_{\rm final}$ gives the redshift that the simulation has been
evolved to.

We add that we ran five realisations of B20 to redshift $z$=10 in
order to have a statistically significant sample of halos in that
particular model. However, we also note that the fitting parameters 
presented in the following Section are robust in the sense that they 
do not depend on whether we stack the halos from those five runs or 
use them individually.

\section{The Halo Sample} \label{sec:virialisation}
We show in a companion paper (Power, Knebe \& Knollmann 2008) that a
substantial fraction of the halo population at high redshift is not in
virial equilibrium. Because we wish to examine the spin
distribution of equilibrium halos, it is important to account for
unrelaxed systems when investigating correlations between spin and
halo mass. For example, it has been shown that the spin can increase
sharply in the aftermath of mergers with mass ratios as modest as 5:1
(e.g. Hetznecker \& Burkert 2006), and that the degree to which a halo
is in dynamical relaxation is as important as recent merging history in its
influence on spin (D'Onghia \& Navarro 2007). To ensure
that halos in our sample are in virial equilibrium, we compute the
virial ratio for each halo, which we define as

\begin{equation}
  \label{eq:virial}
  Q = \frac{2T+S_p}{U} + 1 \ .
\end{equation}

\noindent Here $T$ represents the kinetic energy, $U$ the potential energy,
and $S_p$ the surface pressure of a given halo of mass \Mhalo. By
including $S_p$, we can account for the effect of infalling material on
the dynamical state of the halo. Each of these quantities are evaluated
using all gravitationally bound particles, and we adopt the formula of 
Shaw~\ea (2006) for the surface pressure term~$S_p$ (cf. equations.(4)-(6) in 
their study).

\begin{figure}
  \plottwo{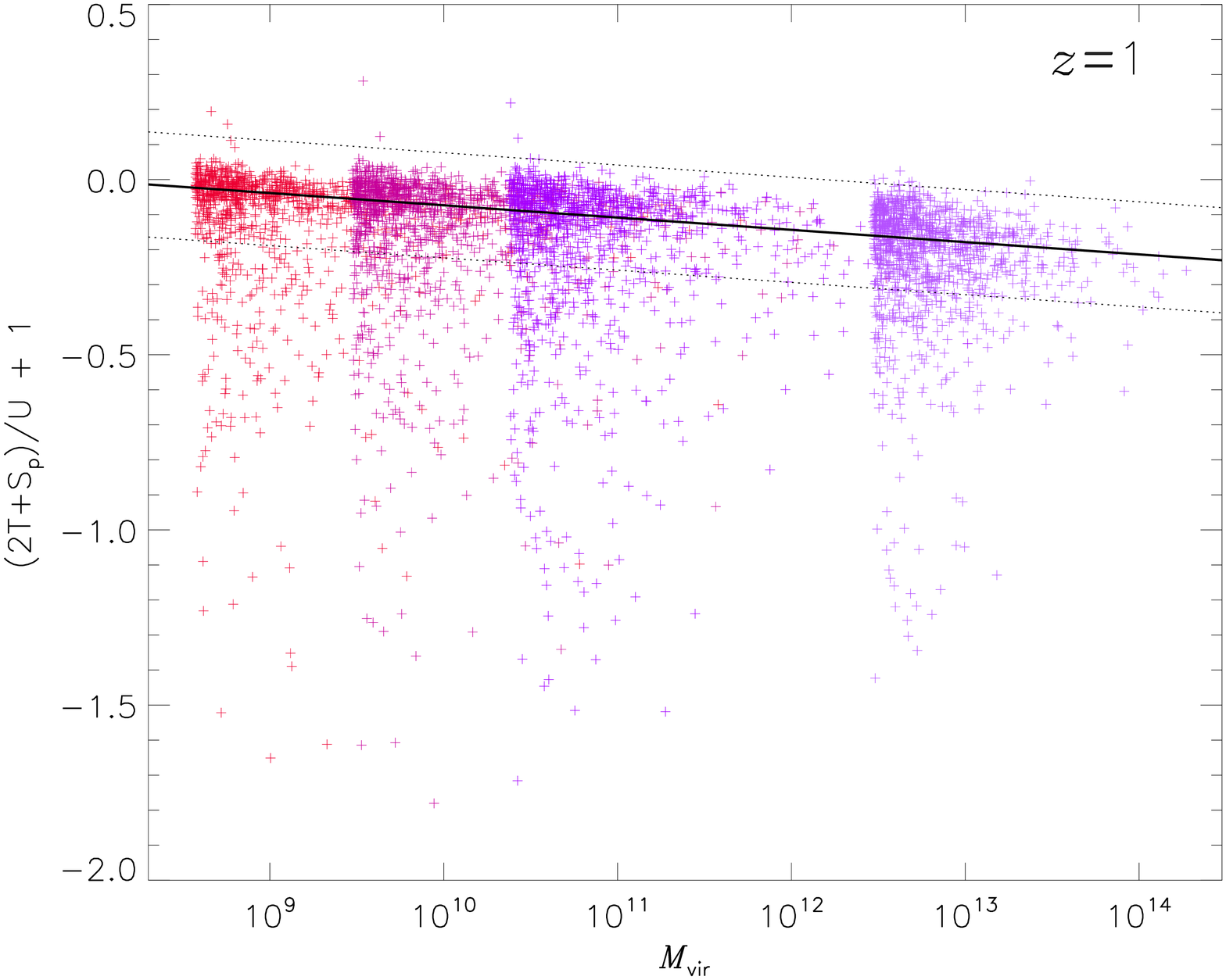}{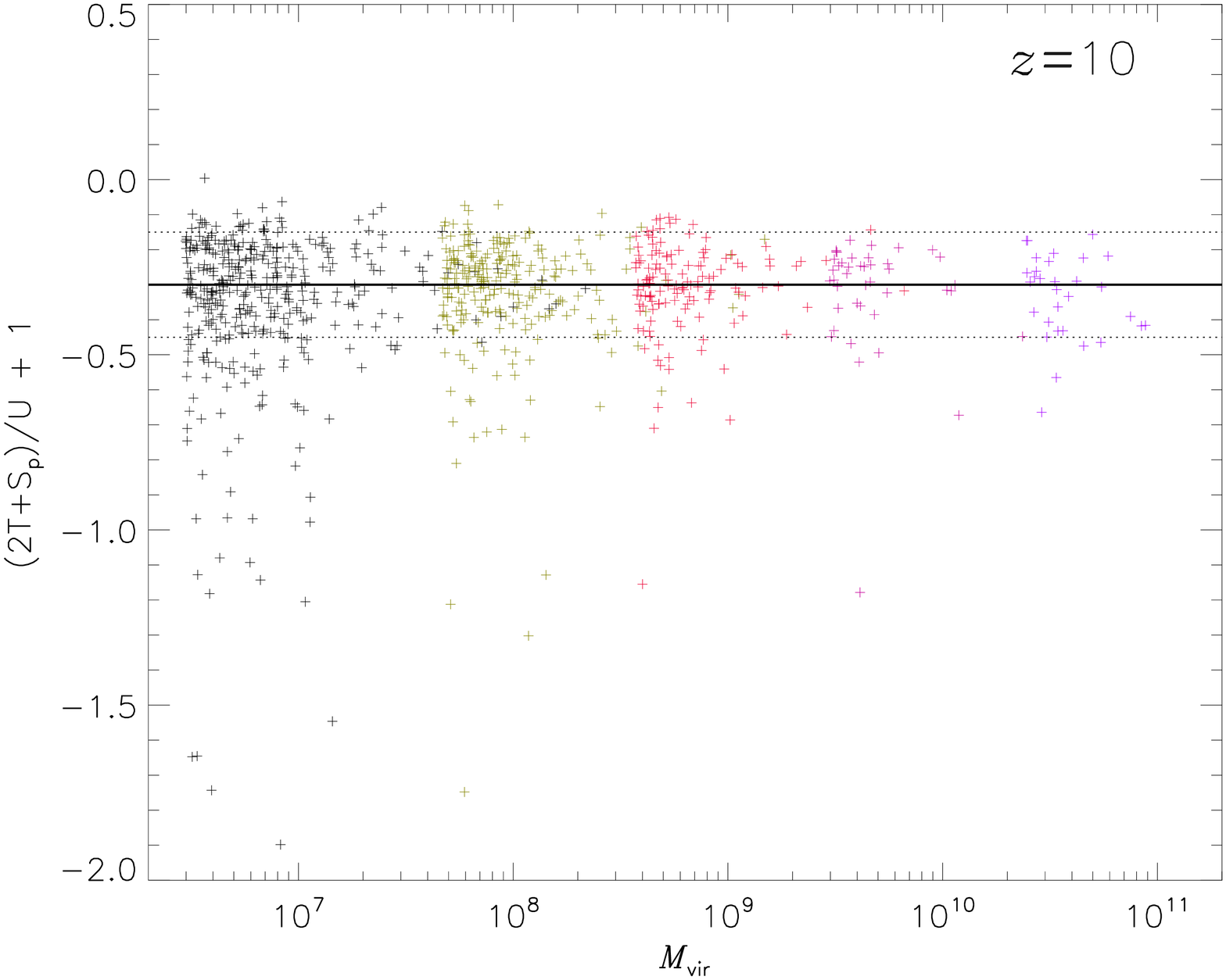}
  \caption{Relation between virial parameter $Q$
    (cf. equation~\ref{eq:virial}) and halo mass.  The solid lines
    represent the adopted virialisation criteria as given by
    \Eq{eq:QspMass}. Note that we already applied the mass cut of 600
    particles per halo for this plot and hence the number of halos appearing
    does not agree with the number given in \Table{tab:simu}.}\label{fig:QspMass}
  \end{figure}

In Figure~\ref{fig:QspMass} we show that the relation between halo
mass and $Q$ can vary with mass. This is apparent at redshift $z$=1, 
where we find a trend for more massive halos to be less virialised. 
In contrast, high redshift halos are less virialised on average 
(as indicated by the increased average $\langle Q \rangle\approx-0.3$), 
but we find no apparent trend with mass.

Why is there a mass dependence at $z$=1 but not at $z$=10? There are
two factors. The first is that high redshift halos ``see'' an effective 
slope of the initial power spectrum of $n_{\rm eff} \approx -3$, and so 
the time at which a particular mass scale starts to collapse is
relatively insensitive to mass. Therefore we do not expect to find a
strong correlation between virial state $Q$ and mass. We have 
checked this halo populations drawn from the simulations of scale-free 
cosmologies of Knollmann, Power \& Knebe (2008) and our interpretation 
is consistent with the correlations we find in these runs.
The second is that the typical collapsing mass $M^{*}$ at $z$=10 is
small -- of order $10^3 h^{-1} \rm M_{\odot}$ -- and because we resolve 
mass scales that have collapsed more or less simultaneously, we see a 
population that has yet to relax. At $z$=1, the typical collapsing mass 
is much larger -- of order $10^{11} h^{-1} \rm M_{\odot}$ -- and so we 
resolve a population of halos whose mass accretion histories are more diverse. 
The most massive systems tend to be ones that have formed most
recently, and are therefore the least dynamically relaxed.

We have used the following relation between $Q$ and $M$ to classify
dynamically relaxed and unrelaxed systems:

\begin{equation} \label{eq:QspMass}
\begin{array}{rcll}
Q_{\rm allowed} & \propto & M^{-0.015}         & \mbox{, for $z=1$}  \\ 
Q_{\rm allowed} & \propto & \mbox{const.}    & \mbox{, for $z=10$} \\ 
\end{array}
\end{equation}

\noindent We allow the $Q$ values of halos in our sample to deviate from these
scaling relations by not more than 

\begin{equation} \label{eq:Qlim}
Q_{\rm allowed}-Q_{\rm lim} \le Q \le Q_{\rm allowed}+Q_{\rm lim}
\end{equation}

\noindent with $Q_{\rm lim} = 0.15$ (indicated by the dashed lines in
\Fig{fig:QspMass}). Furthermore, we consider only halos that contain
at least $N_{\rm min}=600$ particles within their virial radius to
ensure that we are not influenced by particle discreteness. Interestingly, 
when computing spin, the tighest restriction on particle number comes not 
from the calculation of angular momentum but from the calculation of the 
potential energy. By comparing analytic solutions with Monte Carlo 
realisations of Navarro, Frenk \& White (1997) haloes, we find that at least
600 particles are required if the energy is to be computed to better than 
$10\%$.

\section{The Spin-Mass Correlation} \label{sec:correlation}
Calculating the total energy $E$ of a halo is computationally expensive, and so
computing $\lambda$ using \Eq{eq:lambda} is also expensive. This prompted 
Bullock et al. (2001) to introduce a modified spin parameter

\begin{equation} \label{eq:spinTsp}
  \lambda' = \frac{J}{\sqrt{2}MVR} \ ,
\end{equation}

\noindent where $V=\sqrt{GM/R}$ measured the circular velocity at the
virial radius $R$ and $J$ represents the absolute value of the angular
momentum. We follow Bullock et al. and compute spin using
equation~(\ref{eq:spinTsp}).

\begin{figure}
\plottwo{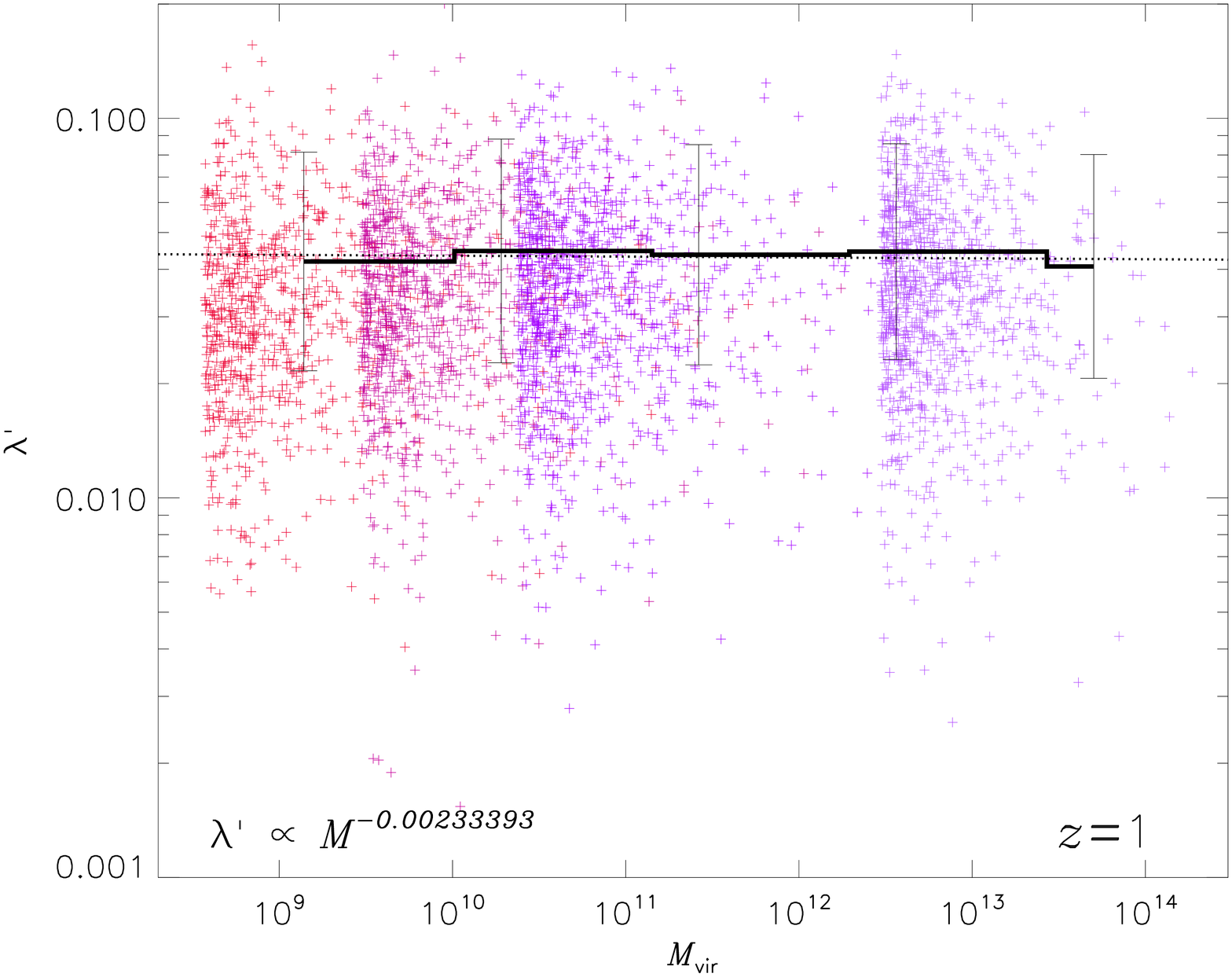}{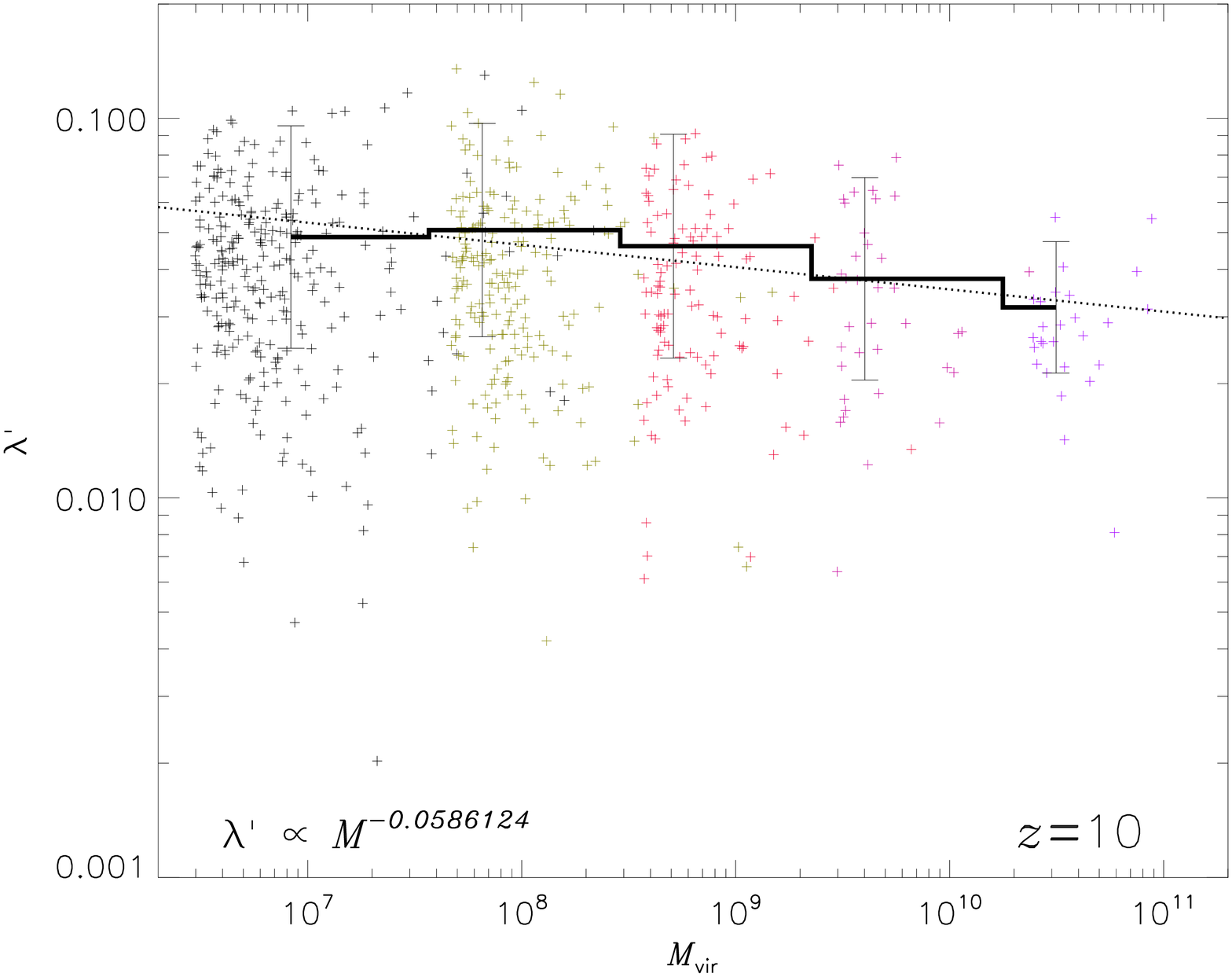}
\caption{Correlation of spin parameter $\lambda$ with mass $M$. The
  binned data (histograms) has been fitted to a power laws (dashed
  line, cf. \Eq{eq:logslopes}).
         \label{fig:SpinMass}}
\end{figure}

In Figure~\ref{fig:SpinMass}, we investigate the correlation between
halo spin $\lambda'$ and mass $M$. We show only halos that fulfill our
selection criteria (individual dots) and bin the data in five mass
bins equally spaced in log-space between $M_{\rm min}$ and $M_{\rm
  max}$ of the considered halos at the respective redshift. The values
plotted as histograms thereby represent the weighted mean of all spin
parameters in the respective mass range where the weight is inversely
proportional to the error estimate

\begin{equation}
 \frac{\sigma_J}{J} = \frac{0.2}{\lambda' \sqrt{N}}
\end{equation}

\noindent
for the spin parameter of a halo consisting of $N$ particles as
derived in Bullock et al. (2001) (cf. equation~(7) in that study). The
error bars indicate the standard deviation of the spin parameter
values in the bin from the weighted mean.\footnote{We like to note in
  passing that we also performed all of the analysis and stability
  checks using the median and the scatter about the median in each
  bin. The results remain unaffected and we therefore decide to only
  list them for the weighted means.}

The best fitting power-laws to these histograms reveal that

\begin{equation}\label{eq:logfit}
\begin{array}{lcll}
\displaystyle \lambda'  & \propto & M^{\alpha} \\ 
\end{array}
\end{equation}

\noindent with

\begin{equation}\label{eq:logslopes}
\begin{array}{lcll}
\displaystyle \alpha  & = & -0.002 \pm 0.149 & \mbox{\rm ,\ for $z=1$ } \\
\displaystyle \alpha  & = & -0.059 \pm 0.171 & \mbox{\rm ,\ for $z=10$ } .\\ 
\end{array}
\end{equation}

This indicates that there is a \emph{weak} correlation at high
redshifts for spin to decrease with increasing mass, albeit stronger than the
one at $z$=1. We compute Spearman rank correlation coefficients at $z$=10 (1) 
and find $R_s=-0.137 (0.06)$.

As an alternative approach, we fit a lognormal function to each of
our halo samples at $z$=10 and $z$=1,

\begin{equation} \label{lognormal}
 P(\lambda') = \displaystyle \frac{1}{\lambda' \sqrt{2\pi} \ \sigma_0}
              \exp \left( {-\frac{\ln^2 (\lambda'/\lambda'_0)}{2 \sigma_0^2}} \right).
\end{equation}

\begin{figure}
\plottwo{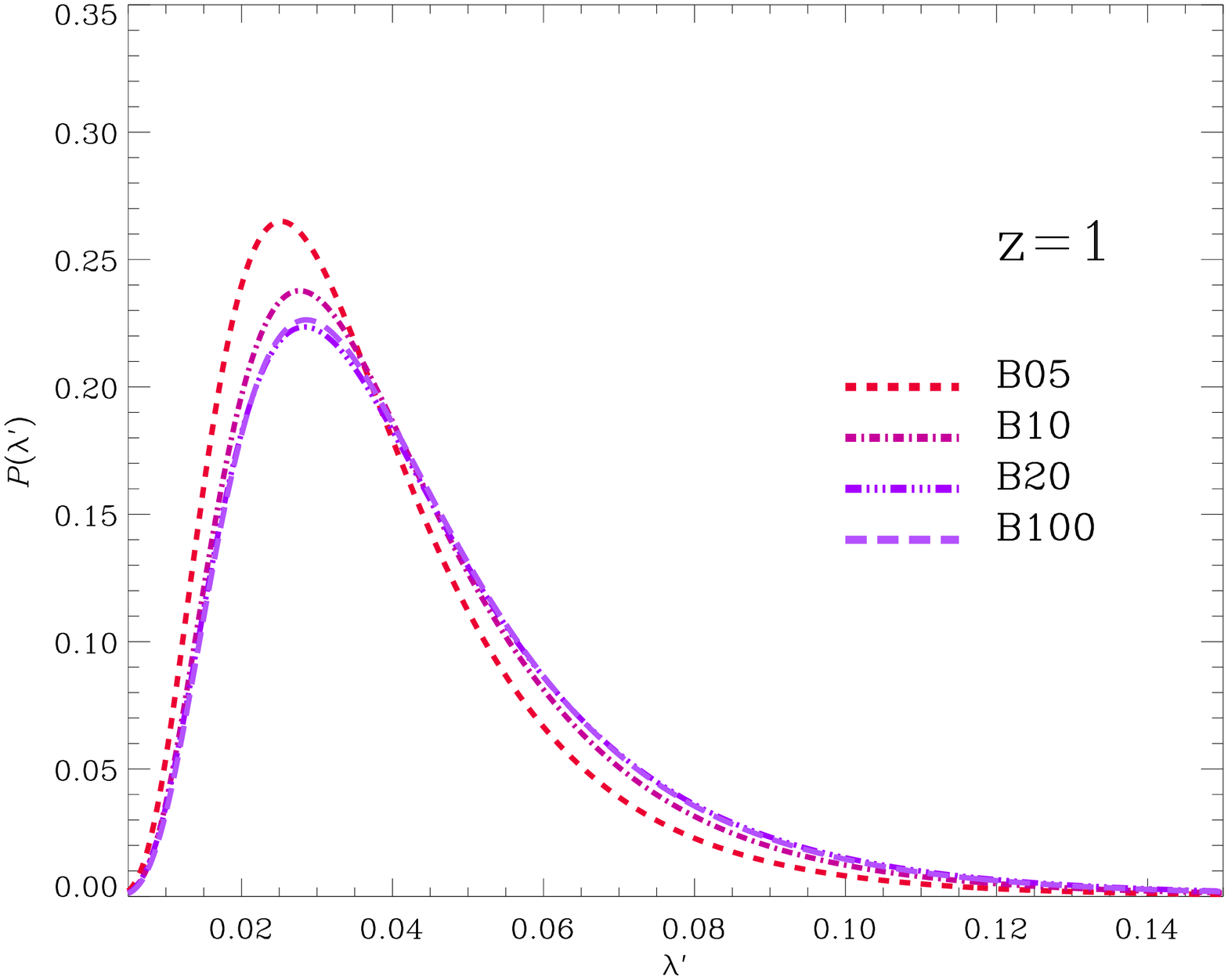}{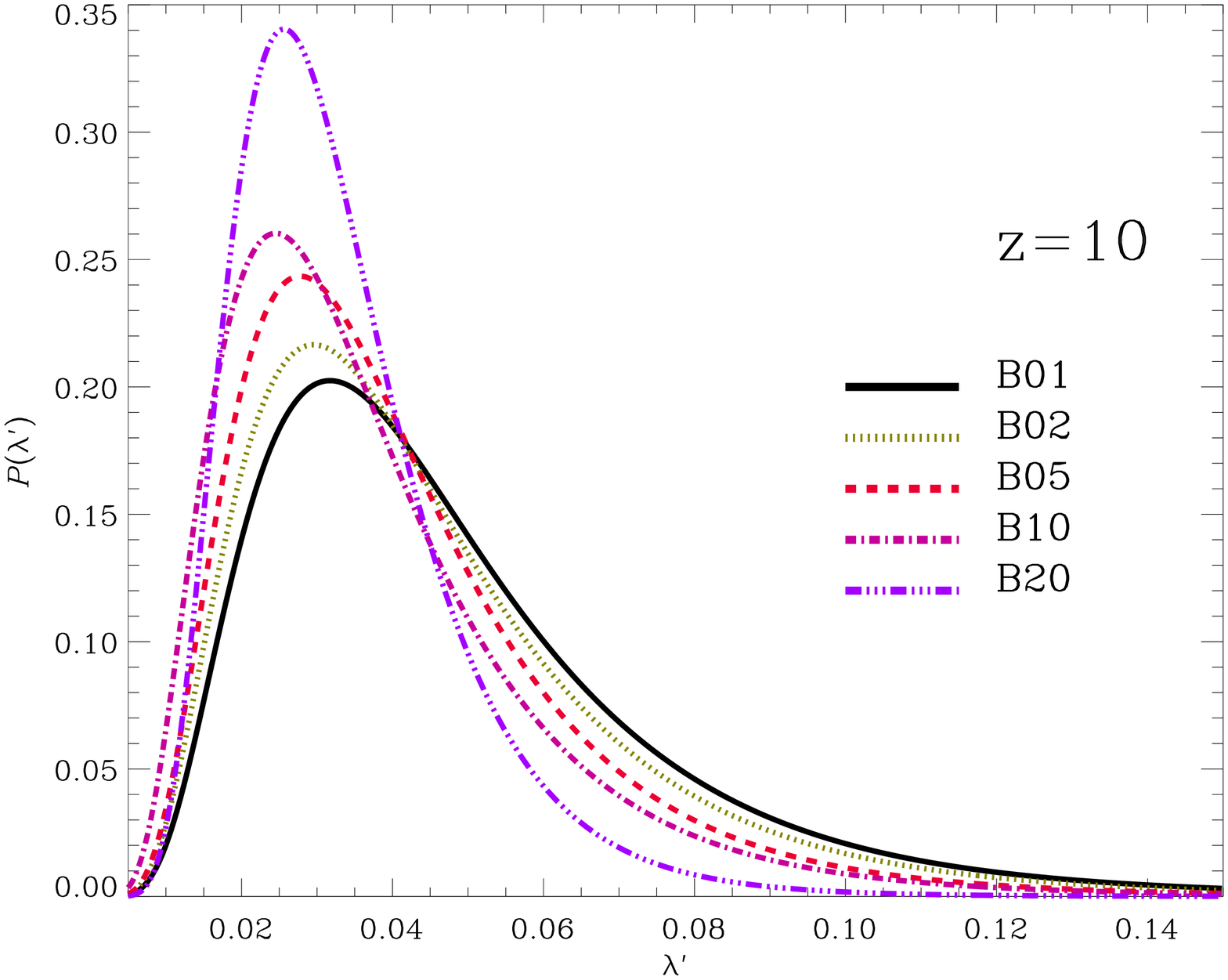}
\caption{Lognormal distributions of the spin parameter $\lambda'$. 
         \label{fig:Pspin}}
\end{figure}

\noindent The resulting curves are presented in \Fig{fig:Pspin}
whereas the best-fit parameters, median values for $\lambda'_{\rm med}
= {\rm median}(\lambda')$ and median halo masses $M_{\rm med}$ are
given in \Table{tab:lambdaTsp}. Inspection of the best-fit parameters
confirm that the median spin declines as we move from less massive to
more massive objects at high redshift.

\section{Stability of Results} \label{sec:stability}
Because of the weak nature of the measured correlation it is
vitally important to check its credibility by performing a statistical
analysis. To this extent we investigate the sensitivity of the
logarithmic slope $\alpha$ with respect to a number of parameters that
enter into its determination, namely the number of bins $N_{\rm bins}$ 
used for the histograms; the virialisation criterion
parametrized via $Q_{\rm lim}$; and the minimum number of particles
$N_{\rm min}$ within a halo's virial radius. Note that we vary one
parameter at a time, keeping the others at their fixed ``standard'' values.  
The results are presented in \Table{tab:Nbins}, \Table{tab:Qlim}, and
\ref{tab:Nmin}.

We find that bin number has practically no effect on the
slope. Similarly we find that varying the virialisation criterion
$Q_{\rm lim}$ has little effect on the slope of the relation between
mass and spin, regardless of redshift (\Table{tab:Qlim}). In contrast,
we find that the minimum number of particles within a halo's virial
radius has a strong and systematic effect on the result at $z$=1 -- as
$N_{\rm min}$ increases, we find that the logarithmic slope becomes
shallower. This does not appear to be true at high redshift, although
as we go to earlier times we find that the number of massive haloes
becomes progressively smaller and our determination of $\alpha$
becomes increasingly unreliable.

These tests lead us to believe that our results are both stable and reliable, 
and our main result holds: \emph{the correlation between spin paramater 
$\lambda'$ and halo mass $M$ is one order of magnitude larger at redshift
$z$=10 than at $z$=1.}\\

As a further test of the credibility of the correlation we measure between 
halo mass and spin, we use the criteria of three other studies to select our
 halo sample. These are:

\begin{itemize}

\item Maccio et al. (2007) criteria:
  
  \begin{itemize}
  \item $N_{\rm min}=250$
  \item $x_{\rm off} < 0.04$
  \item $\rho_{\rm rms} < 0.4$
  \end{itemize}

\item Bett et al. (2007) criteria:
  
  \begin{itemize}
  \item $N_{\rm min}=300$
  \item $Q_{\rm lim}=0.5$
  \end{itemize}

\item Neto et al. (2007) criteria:
  
  \begin{itemize}
  \item $N_{\rm min}=600$
  \item $x_{\rm off} < 0.07$
  \item $f_{\rm sub} < 0.1$
  \end{itemize}

\end{itemize}

Here $N_{\rm min}$ is again the minimum number of particles in a halo,
$Q_{\rm lim}$ the virial limit as defined in \Eq{eq:Qlim}, $x_{\rm
  off}$ measures the distance between the most bound particle and the
centre of mass in units of the virial radius, $\rho_{\rm rms}$ is an
indicator of how well the density profile of the halo can be fitted by
a Navarro, Frenk~\& White (1997) profile (cf. Eq.~(2) in Maccio et
al. 2007), and $f_{\rm sub}$ is the fraction of mass in subhalos. The
resulting power law slopes $\alpha$ (cf. \Eq{eq:logfit}) for $z=1$ and
$z=10$ are presented in \Table{tab:EtAl}. Again we note that there
appears to be a much stronger correlation between $\lambda'$ and $M$
at higher redshift. While the relation is consistent with zero at
$z$=1 (as confirmed by Maccio et al. 2007) spin and mass are
correlated at $z$=10. How this result relates to the Bett et al. (2007) 
result -- who find a weak correlation at $z$=0 -- will be discussed in the
following Section.

\section{Conclusions and Discussion} \label{sec:conclusions}
We have performed a careful investigation of the relation between
virial mass and dimensionless spin parameter for dark matter halos
forming at high redshifts $z \gtrsim 10$ in a $\Lambda$CDM cosmology.
The result of our study, which is based on a series of cosmological
$N$-body simulations in which box size was varied while keeping
particle number fixed, indicates that there is a \emph{weak}
correlation between mass and spin at $z$=10, such that the spin
decreases with increasing mass. If there is a correlation
at $z$=1, we argue that it is significantly weaker than the one we find
at $z$=10; this is in qualitative agreement with the findings of
previous studies that focused on lower redshifts 
(Maccio~\ea 2007, Shaw~\ea 2005, Lemson~\& Kauffmann 1999).\\ 

Interestingly, B07 find a weak correlation between median spin and
halo mass at $z$=0 in the Millennium Simulation (Springel et
al. 2005), in the same spirit as the one presented here for $z$=10:
lower mass halos tend to have higher spins. However, as we show, the
correlation between halo mass and spin is weaker at $z$=1 than at
$z$=10, whereas the correlation reported in B07 for halos at $z$=0 is
much stronger than the one we find at $z$=1. This is not what one
would expect, and so it is important to try and understand the source
of the difference between our result at $z$=1 and the B07 result at
$z$=0. B07 fitted a 3$^{\rm rd}$-order polynomial to the median spins
of halos in the mass range $3\times 10^{11} \lesssim M/(h^{-1}
M_{\odot}) \lesssim 3\times 10^{14}$ at $z$=0. The form of this
polynomial is extremely sensitive to the precise values of the
best-fit parameters (Bett, private communication) and it is not
straightforward to extrapolate its behaviour outside of the given mass
range and redshift. We derive our estimates of the power-law exponents
from the spin distribution with respect to halo mass at $z$=1. Our
halos lie in the mass range $3\times 10^{9} \lesssim M/(h^{-1}
M_{\odot}) \lesssim 5\times 10^{12}$. B07 base their median spins upon
$\sim 1.5$ million halos with correspondingly small errors, and note
that the weak nature of the trend of spin with mass makes it hard to
detect. This suggests to us that the correlation between mass and spin
at $z$=10 is remarkably strong rather than the correlation at $z$=1
being too weak!

When studying correlations between halo mass and spin, great care must
be taken in defining the halo sample. In particular, we find that mass
resolution (i.e. the number of particles with which a halo is
resolved) and the degree of virialisation of a halo can have a
significant effect on the strength of the correlation (at least at
$z$=1, cf. \Table{tab:Nmin}). This -- at least -- is in good agreement
with the findings of B07.\\

We note that Power \& Knebe (2006) demonstrated that the size of 
simulation box can lead to a suppression of angular momentum in 
smaller boxes, due to the absence of longer wavelength perturbations 
in the initial conditions. This will lead to a bias in our estimate of
$\lambda$ (approximately a $\sim 10\%$ effect) but we have verified
that the spin distributions we obtain from a B20 run truncated on
scales larger than the longest wavelength perturbation modelled in the
B1 run produces results that are consistent. Indeed, we would expect
the correlation to be strengthened if the B1 spins were corrected for
box size effects.\\

It is interesting to speculate on the consequences of this correlation for
galaxy formation at high redshifts and the galaxy population we
observe today. In the standard picture of galaxy formation, gas cools
on to dark matter halos and is shock heated to the virial temperature
of the halo. The angular momentum of the gas and the dark matter should
(initially) be similar because they are subject to the same tidal
field. As the innermost densest parts of the gaseous halo cool, they
will settle into a gaseous disk with a scale length determined by the
specific angular momentum of the gas, which we would expect to be
related to the angular momentum of the halo (e.g. Zavala, Okamoto \&
Frenk 2007).

If more massive halos at high redshifts show a tendency to have smaller spin
parameters, the gas disks will have lower specific angular momenta and 
therefore will be more centrally concentrated. If star formation rate 
correlates with surface density, then we might expect the 
star formation rate to be enhanced in more massive halos. Because massive halos
tend to form preferentially in high density, highly clustered environments in 
which the merger rate also tends to be enhanced, then we might expect star
formation to proceed more rapidly and at earlier times in these halos.
Might this explain the effect of ``downsizing'' (e.g. Cowie et al. 1996), 
the successive shifting of star formation from high- to low-mass galaxies 
with decreasing redshift? We shall pursue this in a more quantitative manner 
in future work.

\acknowledgments
AK would like to thank Swinburne University for its hospitality where
this work was initiated. AK further acknowledges funding through the
Emmy Noether programme of the DFG (KN 755/1). CP acknowledges funding
through the ARC Discovery Projected funded ``Commonwealth Cosmology
Initiative'', grant DP 0665574. The simulations presented in this
paper were carried out on the Beowulf cluster at the Centre for
Astrophysics~\& Supercomputing, Swinburne University as well as the
Sanssouci cluster at the Astrophysikalisches Institut Potsdam.

\clearpage

\begin{deluxetable}{ccccccccc}
\tablecaption{Summary of the cosmological simulations and number of halos. \label{tab:simu}}
\tablewidth{0pt}
\tablehead{ 
\colhead{run}                             & 
\colhead{$L_{\rm box}$ [\hMpc]}                 & 
\colhead{$m_p$ [\hMsun]}                   & 
\colhead{$z_{\rm final}$}                    & 
\colhead{$N_{\rm halos}^{z=1}$}     &
\colhead{$N_{\rm halos}^{z=10}$}  &
\colhead{$N_{\rm relaxed\ halos}^{z=1}$}     &
\colhead{$N_{\rm relaxed\ halos}^{z=10}$} 
}
\startdata
 B01 &  1\ \ \ & 4.9 $\times 10^3$ & 10 &    ---   &  8780  &  ---   & 286  \\
 B02 &  2.5    & 7.8 $\times 10^4$ & 10 &    ---   &  7991   &   ---   & 201  \\
 B05 &  5\ \ \ & 6.2 $\times 10^5$ & 1  &   16917  &  6532 &  832    & 109 \\
 B10 &  10     & 4.9 $\times 10^6$ & 1  &   18589  &  4360  &  949    & 37 \\
 B20 &  20     & 4.0 $\times 10^7$ & 1  &   20514  &  10947 &  995    & 27  \\
B100 &  100    & 4.9 $\times 10^9$ & 1  &   24696  & ---    &  978    & ---\\
\enddata
\end{deluxetable}

\begin{deluxetable}{ccccccccccc}
\tablecaption{Fitting parameters for $P(\lambda)$. \label{tab:lambdaTsp}}
\tablewidth{0pt}
\tablehead{ 
\colhead{run}                        & 
\colhead{$\lambda_0'$}                & 
\colhead{$\sigma$}                   & 
\colhead{$\lambda_{\rm med}'$}         & 
\colhead{$M_{\rm med}$ [\hMsun]}      &
&
\colhead{run}                        & 
\colhead{$\lambda_0'$}                & 
\colhead{$\sigma$}                   & 
\colhead{$\lambda_{\rm med}'$}         & 
\colhead{$M_{\rm med}$ [\hMsun]}      
}
\startdata
\multicolumn{5}{c}{$z$=1}  & & \multicolumn{5}{c}{$z$=10} \\
\hline
 \multicolumn{5}{c}{---}                & & B01 & 0.042 & 0.538 & 0.041 & 5.95e+06 \\
 \multicolumn{5}{c}{---}                & & B02 & 0.040 & 0.539 & 0.037 & 8.19e+07 \\
B05  & 0.033 & 0.521 & 0.031 & 7.92e+08 & & B05 & 0.036 & 0.516 & 0.035 & 5.63e+08 \\
B10  & 0.037 & 0.527 & 0.034 & 6.15e+09 & & B10 & 0.033 & 0.540 & 0.029 & 4.14e+09 \\
B20  & 0.038 & 0.543 & 0.036 & 5.40e+10 & & B20 & 0.030 & 0.251 & 0.027 & 3.33e+10 \\
B100 & 0.037 & 0.535 & 0.035 & 5.25e+12 & &       \multicolumn{5}{c}{---}       \\
\enddata
\end{deluxetable}

\begin{deluxetable}{ccccc}
\tablecaption{Variation of $\alpha$ with $N_{\rm bins}$ ($Q_{\rm lim}=0.15, N_{\rm min}=600$). \label{tab:Nbins}}
\tablewidth{0pt}
\tablehead{ 
\colhead{$N_{\rm bins}$}     & 
\colhead{$\alpha_{z=1} \pm \sigma_\alpha$}   & 
\colhead{$\alpha_{z=10} \pm \sigma_\alpha$}
}
\startdata
4            & -0.001 $\pm$ 0.169 & -0.061 $\pm$ 0.191  \\
5            & -0.002 $\pm$ 0.149 & -0.059 $\pm$ 0.171 \\
6            & -0.005 $\pm$ 0.137 & -0.053 $\pm$ 0.155  \\
7            & -0.006 $\pm$ 0.128 & -0.056 $\pm$ 0.144  \\
8            & -0.003 $\pm$ 0.120 & -0.069 $\pm$ 0.132  \\
\enddata
\end{deluxetable}

\begin{deluxetable}{ccccc}
\tablecaption{Variation of $\alpha$ with $Q_{\rm lim}$ ($N_{\rm bins}=5, N_{\rm min}=600$). \label{tab:Qlim}}
\tablewidth{0pt}
\tablehead{ 
\colhead{$Q_{\rm lim}$}     & 
\colhead{$\alpha_{z=1} \pm \sigma_\alpha$}   & 
\colhead{$\alpha_{z=10} \pm \sigma_\alpha$}  &
\colhead{$N_{\rm relaxed\ halos}^{z=1}$}   & 
\colhead{$N_{\rm relaxed\ halos}^{z=10}$}
}
\startdata
0.05 & -0.007 $\pm$ 0.148 & -0.058 $\pm$ 0.153 & 2123 & 264 \\
0.10 & -0.004 $\pm$ 0.148 & -0.062 $\pm$ 0.168 & 3365 & 497 \\
0.15 & -0.002 $\pm$ 0.149 & -0.059 $\pm$ 0.171 & 3811 & 660 \\
0.20 & -0.005 $\pm$ 0.151 & -0.052 $\pm$ 0.173 & 3994 & 739 \\
0.25 & -0.002 $\pm$ 0.153 & -0.052 $\pm$ 0.173 & 4165 & 774 \\
\enddata
\end{deluxetable}

\begin{deluxetable}{ccccc}
\tablecaption{Variation of $\alpha$ with $N_{\rm min}$ ($Q_{\rm lim}=0.15, N_{\rm bins}=5$). \label{tab:Nmin}}
\tablewidth{0pt}
\tablehead{ 
\colhead{$N_{\rm min}$}     & 
\colhead{$\alpha_{z=1} \pm \sigma_\alpha$}   & 
\colhead{$\alpha_{z=10} \pm \sigma_\alpha$}  &
\colhead{$N_{\rm relaxed\ halos}^{z=1}$}   & 
\colhead{$N_{\rm relaxed\ halos}^{z=10}$}
}
\startdata
100  &  0.003 $\pm$ 0.129 & -0.037 $\pm$ 0.158 & 19707 & 5478 \\
200  &  0.002 $\pm$ 0.137 & -0.038 $\pm$ 0.167 & 10809 & 2534 \\
300  &  0.002 $\pm$ 0.142 & -0.035 $\pm$ 0.174 & 7336  & 1526 \\
600  & -0.002 $\pm$ 0.149 & -0.059 $\pm$ 0.171 & 3811  &  660 \\
1000 & -0.004 $\pm$ 0.155 & -0.058 $\pm$ 0.196 & 2303  &  343 \\
2000 & -0.011 $\pm$ 0.164 & -0.048 $\pm$ 0.186 & 1154  &  120 \\
\enddata
\end{deluxetable}

\begin{deluxetable}{ccccc}
\tablecaption{Applying different virialisation criterion ($N_{\rm bins}=5$). \label{tab:EtAl}}
\tablewidth{0pt}
\tablehead{ 
\colhead{criterion}   & 
\colhead{$\alpha_{z=1}$}   & 
\colhead{$\alpha_{z=10}$}  &
\colhead{$N_{\rm relaxed\ halos}^{z=1}$}   & 
\colhead{$N_{\rm relaxed\ halos}^{z=10}$}
}
\startdata
Neto et al. (2007)   &  0.000 $\pm$ 0.149 & -0.041 $\pm$ 0.178 & 3486 & 429 \\
Maccio et al. (2007) &  0.001 $\pm$ 0.146 & -0.040 $\pm$ 0.184 & 4275 & 512  \\
Bett et al. (2007)   &  0.003 $\pm$ 0.146 & -0.040 $\pm$ 0.175 & 8582 & 1955 \\
\enddata
\end{deluxetable}

\clearpage

\end{document}